\documentclass[letterpaper,final]{appolb}

    \usepackage{latexsym,bm,amsmath,amssymb,amsfonts,slashed}
    \usepackage{epsfig,graphics,graphicx}
    \usepackage{cite}

    \newcommand{\dif}{{\rm d}}
    \newcommand{\abar}{\bar{\alpha}}

    \newcommand{\rme}{{\rm e}}
    \newcommand{\rmi}{{\rm i}}
    
    \newcommand{\Lam}{\Lambda_{{\rm QCD}}}

    \newcommand{\mcal}{\mathcal}

    \newcommand{\beq}{\begin{eqnarray}}
    \newcommand{\eeq}{\end{eqnarray}}

\begin{document}

\title{FORWARD AND MUELLER-NAVELET JETS\thanks{Presented at ``Excited QCD 2011'', Les Houches, France, February 2011 [To appear in Acta Physica Polonica B Proceedings Supplement].}}
\author{D.N.~Triantafyllopoulos
\address{ECT*, Strada delle Tabarelle 286, I-38123, Villazzano (TN), Italy}}
\maketitle

\begin{abstract}
We discuss the production of forward jets in high energy processes where one probes a dense hadronic wavefunction. In particular, and as a signature of parton saturation, we discuss the possibility of a strong momentum decorrelation in Mueller-Navelet jets which leads to a geometric scaling behavior.
\end{abstract}
\PACS{11.15.Kc, 12.38.Cy}

\section{Introduction}\label{Sect:Intro}

In high energy collisions of two hadrons, a large number of the produced particles emerge in the forward directions. If such a particle of energy $E$ forms an angle $\theta$ with the beam axis, then it is convenient to decompose
its 3-momentum in terms of its rapidity, representing the longitudinal momentum, and its transverse momentum. For massless particles the rapidity is equal to the pseudorapidity $\eta = -\ln \tan(\theta/2)$, while the magnitude of the transverse momentum is $k_{\perp} = E \sin\theta$. Since the forward directions correspond to $\theta$ being close to 0 or $\pi$, the standard belief until a few years ago was that only soft physics is involved and one would have to resort on non-perturbative phenomenological descriptions. However, it is obvious that a large energy of the produced particle can compensate for the smallness of $\theta$, thus rendering $k_{\perp}$ a perturbative scale. For instance, in p-p collisions at the LHC it could be possible to measure jets at forward rapidities up to $|\eta| \simeq 6.5$ and transverse momenta as low as $k_{\perp}\simeq$ 20 GeV.

In such a kinematic regime it is clear that one can use weak coupling techniques. But the largeness of the rapidity indicates that fixed order perturbation theory might fail and calls for possible resummations to obtain more reliable results. Indeed, this is what sometimes is called the semihard region of QCD, where $k_{\perp}$ is much larger than $\Lambda_{\rm QCD}$ but at the same time much smaller than the total energy $\sqrt{s}$ of the process and large logarithms of the type $\ln s/k_{\perp}^2$ appear and need to be resummed.

\begin{figure}
\begin{minipage}[b]{0.48\linewidth}
\begin{center}
\includegraphics[width=0.75\textwidth]{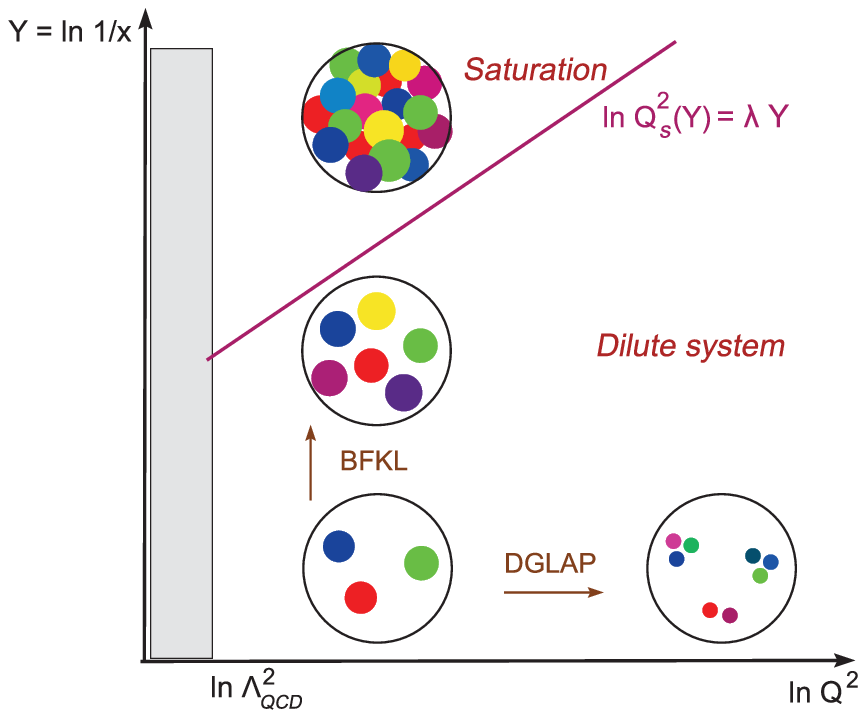}
\caption{\sl Parton evolution in QCD.}
\label{Fig:Evolution}
\end{center}
\end{minipage}
\hspace*{0.02\textwidth}
\begin{minipage}[b]{0.48\linewidth}
\begin{center}
\includegraphics[width=0.75\textwidth]{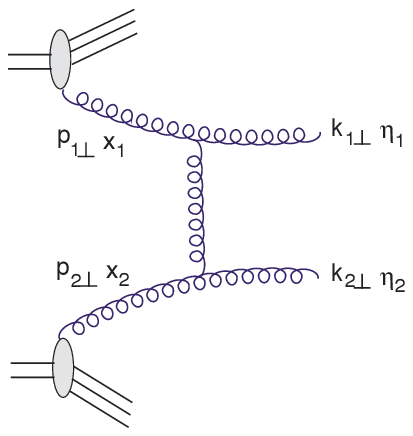}
\caption{\sl Two jets at lowest order.}
\label{Fig:Lowest}
\end{center}
\end{minipage}
\vspace*{-0.3cm}
\end{figure}

Still, the aforementioned resummation might not be enough since when looking at forward directions one is sensitive to the the softer, small-$x$, components of the wavefunction of the one of the incoming hadrons. Here by components we refer to quarks and gluons, with the latter dominating because of their vectorial nature and the 3-gluon vertex in QCD. In the high energy limit and/or for large nuclei (due to an $\sim A^{1/3}$ enhancement coming from the 2D density of nucleons), the hadronic wavefunction becomes dense, partons with sufficiently low, but perturbative, $k_{\perp}$ can overlap and interact with each other leading to a state of saturation \cite{Gelis:2010nm}. Technically this means that one has to perform a resummation of logarithms in the presence of a background field which in general can be strong. This leads to nonlinear equations for the evolution of the partonic densities, with the non-linear terms clearly corresponding to this partonic overlap and becoming important in the saturation regime. Pictorially this is shown in Fig.~\ref{Fig:Evolution} which is our standard picture for parton evolution in QCD. DGLAP is the evolution corresponding to an increase in transverse momentum, and even though it can lead to an increase in the number of partons, these become smaller and smaller due to the uncertainty relation, and the wavefunction remains dilute. BFKL is the evolution corresponding to an increase in $\ln(1/x)$, the number of partons again increases but now they are typically of the same size and the wavefunction becomes denser. For a given  $k_{\perp}$ we reach a point in $x$ where gluons start to feel the presence of each other, and these particular modes saturate reaching their maximal allowed value of order $1/\alpha_s$. Clearly modes with a larger $k_{\perp}$ will saturate at a smaller value of $x$, since the partons will be smaller in size. Therefore one arrives at the concept of the saturation momentum, which increases with $\ln(1/x)$, and which is the borderline between the saturated and the non-saturated modes of the hadronic wavefunction. Some estimates that one can give for the value of the saturation momentum are $Q_s^2 \sim 1$ GeV$^2$ for protons at HERA, $Q_s^2 \sim 1.5$ GeV$^2$ for gold nuclei at RHIC, $Q_s^2 \sim 2$ GeV$^2$ for protons at LHC and $Q_s^2 \sim 6$ GeV$^2$ for lead nuclei at LHC.

\section{Dijet production at lowest order}\label{Sect:Lowest}

At lowest order in perturbation theory the production of two jets is shown in Fig~\ref{Fig:Lowest}. To the order of accuracy the incoming partons, here gluons, participating in the process have zero transverse momentum $\bm{p}_{1\perp}=\bm{p}_{2\perp}=0$. Momentum conservation along this transverse plane implies that $\bm{k}_{1\perp}+\bm{k}_{2\perp} =0$, that is, the azimuthal angle between two outgoing jets is $\Delta\phi=\pi$. Longitudinal momentum and energy conservation constrain the longitudinal momentum fractions $x_1$ and $x_2$ of the incoming partons to be expressed in terms of the center of mass energy $\sqrt{s}$ and the outgoing jets momenta as
 \beq\label{kinematics}
 x_1 = \frac{k_{1\perp}}{\sqrt{s}}\, 
 \rme^{\eta_1} + 
 \frac{k_{2\perp}}{\sqrt{s}}\, \rme^{\eta_2},
 \quad
 x_2 = \frac{k_{1\perp}}{\sqrt{s}}\, 
 \rme^{-\eta_1} + 
 \frac{k_{2\perp}}{\sqrt{s}}\, \rme^{-\eta_2}.
 \eeq
Denoting by PS the phase space $\dif^2\bm{k}_{1\perp} \dif^2\bm{k}_{1\perp} \dif \eta_1 \dif \eta_2$ we have
 \beq\label{LO}
 \frac{\dif \sigma}{\dif \mathrm{PS}} = 
 \sum_{ij}
 x_1 f_{i}(x_1,\mu^2)\, x_2 f_j(x_2,\mu^2)\, 
 \delta^{(2)}(\bm{k}_{1\perp} + \bm{k}_{2\perp})\,
 \frac{\dif \hat{\sigma}_{ij}}{\dif k_{\perp}^2},
 \eeq
where $k_{\perp}$ is the common magnitude of the momenta, $i,j$ refer to the various parton species, $x f_i(x,\mu^2)$ are the corresponding distribution functions and $\hat{\sigma}_{ij}$ the partonic cross sections. The emission of an unobserved extra gluon or quark will smear a bit the peak, but not much, since it is of higher order in perturbation theory and we assume the coupling to be weak.

\section{Two jets in the forward direction}\label{Sect:Forward}

\begin{figure}
\begin{minipage}[b]{0.48\linewidth}
\begin{center}
\includegraphics[width=0.75\textwidth]{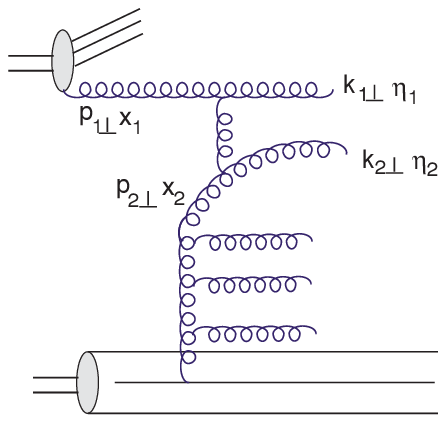}
\caption{\sl Pair of jets in forward direction.}
\label{Fig:Forward}
\end{center}
\end{minipage}
\hspace*{0.02\textwidth}
\begin{minipage}[b]{0.48\linewidth}
\begin{center}
\includegraphics[width=0.75\textwidth]{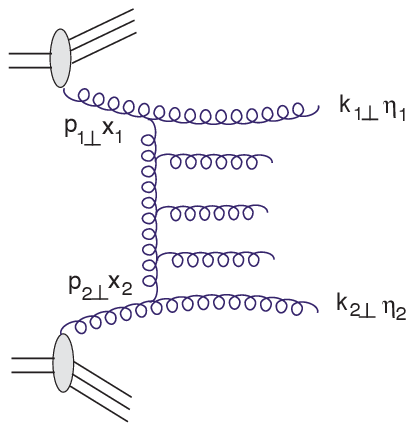}
\caption{\sl Mueller-Navelet jets.}
\label{Fig:MNjets}\end{center}
\end{minipage}
\vspace*{-0.3cm}
\end{figure}

Because of the forward kinematics fixed order perturbation theory is not sufficient. Let us first consider the case where both jets are produced in the 
forward direction as shown in Fig.~\ref{Fig:Forward}, that is we assume that $\eta_1$ and $\eta_2$ are large, say positive, and close to each other. We have in mind collisions of asymmetric objects, like d-Au at RHIC \cite{Albacete:2010pg}, with the forward direction being the one of the colliding deuteron. From the kinematics in Eq.~(\ref{kinematics}) it is straightforward to see that for fixed transverse momenta $x_1$ can be close to 1, but $x_2$ is necessarily very small. The nuclear wavefunction can be dense due to the $A^{1/3}$ enhancement and due to the resummation of large $[\abar_s \ln(1/x_2)]^n$ terms, with $\abar_s = \alpha_s N_c/\pi$ and $N_c$ the number of colors. High energy evolution proceeds via gluon cascades, as shown in Fig.~\ref{Fig:Forward}, which are ordered in longitudinal momentum. Such cascades carry significant transverse momenta, during the scattering they are released and thus there is a large imbalance in the transverse momenta of the two forward jets and the ``away peak'' at $\Delta\phi = \pi$ can disappear. In the BFKL regime one has
 \beq\label{forward}
 \frac{\dif \sigma}{\dif {\rm PS}} = \sum_i x_1 f_i(x_1,\mu^2) \Phi_g(x_2,\bm{k}_{1\perp}+\bm{k}_{2\perp})\,\frac{\alpha_s^2}{k_{1\perp}^4},
 \eeq
where $\Phi_g$ is the unintegrated gluon distribution function whose presence is necessary in order to describe the transverse momentum ``built-up'' along the cascade. Note that Eq.~(\ref{forward}) is valid only so long as the gluon density is not too high and the non-linearities can be neglected. In the presence of saturation the cross-section involves also high point gluon correlators.

\section{Mueller-Navelet jets}\label{Sect:MNjets}

Finally we consider the forward-backward case in p-p (or p-$\bar{\rm p}$) collisions, that is when the two jets are separated by a large rapidity interval. These are the Mueller-Navelet jets \cite{Mueller:1986ey}, where more precisely one has $Y \equiv \eta_1-\eta_2 \gg 1$ with $|\eta_1|\simeq|\eta_2|$ and the corresponding diagram is shown in Fig.~\ref{Fig:MNjets}. We have decided to view the process in a different way than the one of the forward dijet in the previous section. Now the resummation is performed in the partonic cross section, since the emitted gluons in the final state are ordered in rapidity in order to get the dominant contribution. Nevertheless, one can show that the kinematics in Eq.~(\ref{kinematics}) holds approximately and we have $x_1 \sqrt{s} \simeq k_{1\perp} \rme^{\eta_1}$ and $x_2 \sqrt{s} \simeq k_{2\perp}\rme^{-\eta_2}$, so that one can choose $x_1$ and $x_2$ to be ``large'', say around 0.1. We can write the cross section as
 \beq\label{sigmaMNjets}
 \frac{\dif \sigma}{\dif \mathrm{PS}} = 
 \sum_{ij}
 x_1 f_{i}(x_1,\mu^2)\, x_2 f_j(x_2,\mu^2)\, 
 \frac{\dif \hat{\sigma}_{ij}}{\dif^2 \bm{k}_{1\perp} \dif^2 \bm{k}_{2\perp}}.
 \eeq
As explained earlier, at the Born level we get two back-to-back jets and the 
partonic cross section is $\hat{\sigma}_{ij} \sim \mcal{O}(1)$. An extra ``minijet'' introduces some decorrelation in the transverse plane and $\hat{\sigma}_{ij} \sim \mcal{O}(\abar_s Y)$, with two minijets there is more decorrelation and $\hat{\sigma}_{ij} \sim \mcal{O}(\abar_s^2 Y^2)$ and so on. We integrate over the minijets phase space and sum for large $Y$ to find (for any $i,j$)
 \beq\label{sigmaMNpartonic}
 \frac{\dif \hat{\sigma}}{\dif^2 \bm{k}_{1\perp} \dif^2 \bm{k}_{2\perp}} \sim 
 \frac{\alpha_s^2}{k_{1\perp}^2 k_{2\perp}^2}\,
 \Phi(Y,\bm{k}_{1\perp},\bm{k}_{2\perp}),
 \eeq
where $\Phi$ satisfies the BFKL equation. It grows exponential with $Y$ for fixed $\bm{k}_{1\perp}$ and $\bm{k}_{1\perp}$ and imposes transverse momentum decorrelations in the angle and the magnitude \cite{DelDuca:1993mn,Vera:2007kn}\footnote{An NLO BFKL calculation for some observables leads to results that are very close to those obtained from an NLO DGLAP analysis \cite{Colferai:2010wu}.}.

As a particular example of these magnitude decorrelations, let us recall that BFKL dynamics with the necessary inclusion of unitarity corrections leads to geometric scaling \cite{Iancu:2002tr,Mueller:2002zm,Munier:2003vc}, which was introduced and had its greater success in e$^-$p DIS at HERA \cite{Stasto:2000er}. Integrating the transverse momenta above the thresholds $Q_1$ and $Q_2$ we have
 \beq\label{sigmaMNgamma}
 \frac{\dif \sigma}{\dif x_1 \dif x_2} = 
 F_{\rm eff}\,\frac{\alpha_s^2}{Q_2^2}
 \int_{\frac{1}{2} - 
 \rmi \infty}^{\frac{1}{2}+\rmi\infty}\frac{\dif \gamma}{2\pi \rmi} 
 \frac{(Q_2^2/Q_1^2)^{1-\gamma}}{\gamma(1-\gamma)}
 \exp[\abar_s \chi(\gamma)Y]
 +\cdots,
 \eeq
where $F_{\rm eff}$ stands for the contributions from all the quark, antiquark and gluon collinear distributions of the two colliding protons together with some color factors, $\chi(\gamma)$ is the known eigenvalue function of the BFKL equation and the dots stand for unitarity corrections. When $Q_1$ is much larger $Q_2$, but in such a way that the integrand in Eq.~(\ref{sigmaMNgamma}) varies slowly, the integration in the saddle point approximations leads to \cite{Iancu:2008kb}
 \beq\label{sigmaMNscaling}
 \frac{\dif \sigma}{\dif x_1 \dif x_2} = 
 F_{\rm eff}\,\frac{1}{Q_2^2}
 \left( \frac{Q_s^2}{Q_1^2} \right)^{1-\gamma_s}.
 \eeq
In the above $Q_s^2 = Q_2^2 \rme^{\lambda_s(Y-Y_0)}$ is a ``saturation scale'', with $\lambda_s \simeq 0.3$ \cite{Triantafyllopoulos:2002nz}, $Y_0 \sim (1/\abar_s) \ln(1/\alpha_s^2)$ and $\gamma_s=0.372$ is the anomalous dimension dominating the integration. Eq.~(\ref{sigmaMNscaling}) exhibits geometric scaling since, leaving aside the slowly varying prefactor $F_{\rm eff}$, it depends only on two variables, $Q_2^2$ and the combined one $Q_s^2/Q_1^2$, instead of the three $Q_1^2$, $Q_2^2$ and $Y$. This is analogous to the scaling observed in $\gamma^*$p DIS, with the analogy being more precise if we let $Q_2^2 \to \Lam^2$ and $Q_1^2 \to Q^2$, with $Q^2$ the photon virtuality. Thus the softer jet looks like the ``target'' with the harder jet being the ``projectile''.

The difference with DIS is that here we have a very large saturation scale to start with, which is by definition the threshold transverse momentum of the softer jet and which furthermore evolves to higher values with increasing $Y$.  The price to pay is that the cross section is proportional to the ``target'' size squared which is $1/Q_2^2$. The rapidity $Y_0$ corresponds to the amount of evolution which is needed to saturate a small hadronic system, like a dipole or a high-momentum parton, on the resolution scale set by its own size. An NLO estimate gives $Y_0 \simeq 8$ which is very well within the reach of the LHC. 

According to Eq.~(\ref{sigmaMNscaling}), one can explore the scaling behavior of this dijet cross section by keeping the transverse and longitudinal momenta of the softer jet fixed, and vary those of the harder jet so that $x_1$ remains fixed. As said, this represents a strong momentum decorrelation and would test BFKL dynamics and saturation.


\begin{thebibliography}{10}

\bibitem{Gelis:2010nm}
F.~Gelis, E.~Iancu, J.~Jalilian-Marian, and R.~Venugopalan, 
{{\em Ann. Rev. Nucl. Part. Sci.} {\bfseries 60} (2010) 463--489}, 
{{\ttfamily arXiv:1002.0333 [hep-ph]}}.

\bibitem{Albacete:2010pg}
J.~L. Albacete and C.~Marquet, 
{{\em Phys. Rev. Lett.} {\bfseries 105} (2010) 162301}, 
{{\ttfamily arXiv:1005.4065 [hep-ph]}}.

\bibitem{Mueller:1986ey}
A.~H. Mueller and H.~Navelet, 
{{\em Nucl. Phys.} {\bfseries B282} (1987) 727}.

\bibitem{DelDuca:1993mn}
V.~Del~Duca and C.~R. Schmidt, 
{{\em Phys. Rev.} {\bfseries D49} (1994) 4510--4516}, 
{{\ttfamily arXiv:hep-ph/9311290}}.

\bibitem{Vera:2007kn}
A.~Sabio~Vera and F.~Schwennsen, {{\em Nucl. Phys.}
  {\bfseries B776} (2007) 170--186}, {{\ttfamily arXiv:hep-ph/0702158}}.

\bibitem{Colferai:2010wu}
D.~Colferai, F.~Schwennsen, L.~Szymanowski, and S.~Wallon, 
{{\em JHEP} {\bfseries 12} (2010) 026}, 
{{\ttfamily arXiv:1002.1365 [hep-ph]}}.

\bibitem{Iancu:2002tr}
E.~Iancu, K.~Itakura, and L.~McLerran, 
{{\em Nucl. Phys.} {\bfseries A708} (2002) 327--352},
{{\ttfamily arXiv:hep-ph/0203137}}.

\bibitem{Mueller:2002zm}
A.~H. Mueller and D.~N. Triantafyllopoulos, 
{{\em Nucl. Phys.} {\bfseries B640} (2002) 331--350},
{{\ttfamily arXiv:hep-ph/0205167}}.

\bibitem{Munier:2003vc}
S.~Munier and R.~B. Peschanski, 
{{\em Phys. Rev. Lett.} {\bfseries 91} (2003) 232001},
{{\ttfamily arXiv:hep-ph/0309177}}.

\bibitem{Stasto:2000er}
A.~M. Stasto, K.~J. Golec-Biernat, and J.~Kwiecinski, 
{{\em Phys. Rev. Lett.} {\bfseries 86} (2001) 596--599},
{{\ttfamily arXiv:hep-ph/0007192}}.

\bibitem{Iancu:2008kb}
E.~Iancu, M.~S. Kugeratski, and D.~N. Triantafyllopoulos, 
{{\em Nucl. Phys.} {\bfseries A808} (2008) 95--116},
{{\ttfamily arXiv:0802.0343 [hep-ph]}}.

\bibitem{Triantafyllopoulos:2002nz}
D.~N. Triantafyllopoulos, 
{{\em Nucl. Phys.} {\bfseries B648} (2003) 293--316},
{{\ttfamily arXiv:hep-ph/0209121}}.

\end{thebibliography}

\end{document}